\documentclass[letterpaper, 10pt, conference]{ieeeconf}  
\IEEEoverridecommandlockouts                             
\usepackage{graphics}
\usepackage{epsfig} 
\usepackage{amsfonts}
\usepackage{cite}
\usepackage{amsmath}
\usepackage{amssymb}
\usepackage{tcolorbox}
\allowdisplaybreaks[4]
\usepackage{theorem}
\newtheorem{theorem}{Theorem}

\newtheorem{proposition}{Proposition}

\newtheorem{assumption}{Assumption}
\newtheorem{remark}{Remark}
\newtheorem{example}{Example}
\usepackage{subcaption}
\usepackage{comment}
\usepackage{multirow}

\title{\LARGE \bf
A Generalized Stopping Criterion for Real-Time MPC \\
with Guaranteed Stability}

\author{Kristína Fedorová, 
	Yuning Jiang, 
	Juraj Oravec, 
	Colin N. Jones, 
	and Michal Kvasnica
\thanks{KF, JO, and MK gratefully acknowledge the contribution of the Scientific Grant Agency of the Slovak Republic under the grants 1/0490/23, 1/0297/22, and the Slovak Research and Development Agency under the project APVV-20-0261, and the European Commission under grant no. 101079342 (Fostering Opportunities Towards Slovak Excellence in Advanced Control for Smart Industries).
YJ and CNJ gratefully acknowledge the contribution of the Swiss National Science Foundation (SNSF) under the NCCR Automation project, grant agreement 51NF40\_180545.}
\thanks{KF, JO and MK are with the Faculty of Chemical and Food Technology, Slovak University of Technology in Bratislava (STUBA), Slovak. {\tt \{kristina.fedorova, juraj.oravec, michal.kvasnica\}@stuba.sk}}%
\thanks{YJ and CNJ are with the Automatic Control Laboratory, EPFL, Switzerland.
{\tt \{yuning.jiang, colin.jones\}@epfl.ch}}%
}

\begin{document}
\maketitle
\setlength\abovedisplayskip{2.5pt}
\setlength\belowdisplayskip{2.5pt}


\begin{abstract}
Most of the real-time implementations of the stabilizing optimal control actions suffer from the necessity to provide high computational effort. This paper presents a cutting-edge approach for real-time evaluation of linear-quadratic model predictive control (MPC) that employs a novel generalized stopping criterion, achieving asymptotic stability in the presence of input constraints. The proposed method evaluates a fixed number of iterations independent of the initial condition, eliminating the necessity for computationally expensive methods. We demonstrate the effectiveness of the introduced technique by its implementation of two widely-used first-order optimization methods: the projected gradient descent method (PGDM) and the alternating directions method of multipliers (ADMM). The numerical simulation confirmed a significantly reduced number of iterations, resulting in suboptimality rates of less than 2\,\%, while the effort reductions exceeded 80\,\%. These results nominate the proposed criterion for an efficient real-time implementation method of MPC controllers.
\end{abstract}

\section{Introduction}

Model Predictive Control (MPC) is an advanced and widely used control strategy that can effectively address many complex control problems in various fields, including process control~\cite{QIN2003}, automotive control~\cite{Hrovat2012}, and robotics~\cite{Schwenzer2021}. 
The MPC framework operates under the paradigm of moving horizon control strategies and is executed at every control step to account for the current state measurement~\cite{rawlings2017model}. 
It builds up a mathematical model of the system to predict its behavior over a future time horizon. Then, it generates an optimal control action by solving an optimization problem subject to constraints. Therefore, MPC can handle complex system dynamics and consider input and output constraints. In practice, deploying MPC into real-time scenarios requires an efficient and reliable optimization approach to deal with the closed-loop optimal control problem~\cite{QIN2003}. To this end, two conventional approaches are used to implement the MPC control policy, parametric (explicit) MPC~\cite{BM02} and implicit (non-explicit) MPC~\cite{rawlings2017model}. 

The real-time implementation of explicit MPC~\cite{BM02} benefits from efficient division-free computation, yielding an optimal control law in the form of a piecewise affine (PWA) function for \textit{a priori} performance certification. Despite successful implementations, see~\cite{PK21}, and the references therein, the memory limitations and hardly-tractable construction of explicit MPC lead to the use of implicit MPC as an alternative.


Implicit MPC employs iterative methods like active set~\cite{bemporad2015quadratic}, interior-point~\cite{wright1997}, and first-order methods~\cite{goldstein2014fast,patrinos2013accelerated} for online quadratic programming. Active-set methods are quick for small-medium problems but lack scalability and robustness; interior-point methods suit large problems and robustness but struggle with closed-loop warm-starting. First-order methods~\cite{Beck2017} can be easily scaled up, while its slow convergence limits its application in practice. Fortunately, their iterations are computationally cheap, and their runtimes are contingent upon the chosen stopping criterion. As a result, one can stop the iterations in advance to design a suboptimal real-time MPC scheme.


Numerous publications in this field have already been presented, where the authors have focused on establishing stopping criteria for specific optimization algorithms, ensuring asymptotic stability of suboptimal MPC. The authors of~\cite{McGovern1999} introduced a stopping criterion for an interior-point algorithm called the method of centers (MC). The work~\cite{Giselsson2010} gained a stopping criterion formula for the dual decomposition (DD) algorithm. In~\cite{Graichen2012}, authors have derived a criterion for the fixed-point iteration (FPI) algorithm. The authors of~\cite{Rubagotti2014} have presented a formula determining the iteration count for the dual accelerated gradient projection algorithm (DAGPM). The paper~\cite{Liao-McPherson2022} provides a formula for projected gradient (PGM) and accelerated projected gradient method (APGM). Each of the above-mentioned papers defines a different type of stopping criterion while considering only input constraints in MPC (see Table~\ref{tab:stopping_criterion}). The stopping criterion that guarantees asymptotical stability in the presence of state constraints can be found in~\cite{Leung2021} for the projected gradient method (PGM) and in~\cite{Yang2022} for the fast alternating minimization algorithm (FAMA) through invariant set derivation, tightening techniques and additional assumptions. To the author's knowledge, a universal stopping criterion has not yet been published to apply to various optimization algorithms.

In this paper, we formulate a generalized stopping criterion that preserves the asymptotic stability of a closed-loop system while achieving a comparable control performance to the optimal solution with a significant decrease in the number of iterations. The main contribution of this work is twofold,
    (i) we propose a novel generalized stopping criterion representing a maximum finite number of iterations for any first-order optimization method applied to solve the MPC problem. We prove the guarantee of the asymptotic stability of the closed-loop system using a generalized stopping criterion under mild assumptions at the cost of suboptimal control performance;
    (ii) we adopt the generalized stopping criterion to two widely used first-order optimization algorithms solving the MPC problems (PGDM, ADMM). 
    The numerical case study of the well-known double-integrator benchmark is used to analyze the performance loss associated with a reduced number of iterations, accelerating the real-time evaluation.

\begin{table}[htbp!]
	\centering
	\scriptsize
	\renewcommand{\arraystretch}{1.2}
	\caption{Selective summary of stopping criterion for suboptimal MPC.}
	\label{tab:stopping_criterion}
	\begin{tabular}{|c c c c|} 
		\hline
		\multirow{ 2}{*}{Reference} & \multirow{ 2}{*}{Constraints} &  \multirow{ 2}{*}{Fixed no. iterations} & \multirow{ 2}{*}{Algorithm} \\ 
		& \tiny*possible violation & & \\ [0.8ex] 
		\hline\hline
		\cite{McGovern1999} & input & $\checkmark$ & MC \\ 
		\cite{Giselsson2010} & input & -- & DD \\
		\cite{Graichen2012} & input & $\checkmark$ & FPI \\
		\cite{Rubagotti2014} & input* & $\checkmark$ & DAGPM \\ 
		\cite{Liao-McPherson2022} & input & $\checkmark$ & PGM, APGM \\
		\cite{Leung2021} & input, state & $\checkmark$ & PGM \\
		\cite{Yang2022} & input, state & $\checkmark$ & FAMA \\
		\hline
	\end{tabular}
\end{table}

\textit{Notation:} We denote the set of real numbers by $\mathbb{R}$, the set of $n$-dimensional real-valued vectors by $\mathbb{R}^n$, and the set of $n \times m$-dimensional real-valued matrices by $\mathbb{R}^{n \times m}$. Moreover, we denote the subspace of symmetric matrices in $\mathbb{R}^{n \times n}$ by $\mathbb{S}^n$ and the cone of positive (semi-)definite matrices by $\mathbb{S}_{++}^n(\mathbb{S}_+^n)$. For real-valed vector $x$ and the symmetric positive definite matrix $Q$, $\Vert x \Vert_{Q}^{2} := x^{\top} Q x$.



\section{Preliminaries}
\label{sec:problem_statement}

Consider a linear time-invariant (LTI) system in a discrete-time domain having a form
\begin{equation}
\label{eq:lti_system}
x(t+T_{\mathrm{s}}) = A x(t) + B u(t),
\end{equation}
where $x \in \mathbb{R}^{n_\text{x}}$ is a system state vector, $u \in \mathbb{R}^{n_\text{u}}$ is a vector of control actions, $A\in \mathbb{R}^{n_\text{x}\times n_\text{x}}$ is system matrix, $B \in \mathbb{R}^{n_\text{x}\times n_\text{u}}$ is input matrix, and $T_{\mathrm{s}}$ is the sampling time. The corresponding linear-quadratic MPC problem is given by
\begin{equation}
\label{eq:mpc_design_problem}
\begin{aligned}
    V(x_0) = \min_{x,u}&\;\|x_N\|_P^2+\sum_{k=0}^{N-1}\left(\|x_k\|_Q^2 + \|u_k\|_R^2\right) \\
    \text{s.t.}~ & \left\{
    \begin{aligned}
        x_{k+1} & = Ax_k + Bu_k,\;u_k \in\mathbb U_k,\;\\
        \forall k & \in \{0, \dots, N-1\}\\
        x_{0} & = x(t) ,
    \end{aligned}
    \right.
\end{aligned}
\end{equation}
where the decision variables $x = [x_0^{\top},\dots,x_N^{\top}]^{\top} \in \mathbb{R}^{N n_\text{x}}$ and $u = [u_0^{\top},\dots,u_{N-1}^{\top}]^{\top} \in \mathbb{R}^{N n_\text{u}}$ are the sequences of the predicted system states and control actions, respectively, with $x(t) \in \mathbb{R}^{n_x}$ as a state measurement. In~\eqref{eq:mpc_design_problem}, $V$ denotes the minimized value function, $\mathbb{U}_{k} \subset \mathbb{R}^{n_{\mathrm{u}}}$ is the set of the constrained control inputs, the matrices $Q, P \in \mathbb{R}^{n_\text{x}\times n_\text{x}}$, $R \in \mathbb{R}^{n_\text{u}\times n_\text{u}}$ are given tuning parameters, and $N$ is a finite prediction horizon. 

\begin{assumption}
\label{assumption:1}
We assume that for MPC design problem in~\eqref{eq:mpc_design_problem} hold:
\begin{itemize}
    \item the matrix pair $(A,B)$ is controllable,
    \item the penalty factors $Q$, $P$, and $R$ are symmetric positive definite matrices,
    \item terminal penalty $P$ is computed as a solution of the matrix Riccati equation, i.e.,
    \[
    P=A^\top P A - (A^\top PB)(R+B^\top P B)^{-1}+( B^\top P A) + Q,
    \]
    \item the sets $\mathbb U_k$ of control inputs are convex and closed, containing origin in their strict interiors. 
\end{itemize}
\end{assumption}
Assumption~\ref{assumption:1} enforces the strong convexity of the MPC design problem in~\eqref{eq:mpc_design_problem}, leading to a unique optimal solution. Moreover, under Assumption~\ref{assumption:1}, the feasible solution of the MPC design problem in~\eqref{eq:mpc_design_problem} leads to the asymptotic stability of the closed-loop LTI system in~\eqref{eq:lti_system}. As the state constraints are not considered in~\eqref{eq:mpc_design_problem}, the recursive feasibility is satisfied by design. 

The MPC design problem in~\eqref{eq:mpc_design_problem} has the form of the optimization problem of the quadratic programming (QP) in the general form
\begin{equation}
\label{eq:mpc_problem_qp}
\begin{aligned}
    V(x_0)= \min_{u} & \;J(u;x_0) = \frac{1}{2}\begin{bmatrix}
        u \\
        x_0
    \end{bmatrix}^\top \begin{bmatrix}
        H & S \\ 
        S^\top & D
    \end{bmatrix}
    \begin{bmatrix}
        u \\
        x_0
    \end{bmatrix}\quad \\
    \text{s.t.} & \;\; u \in  \mathbb{U}:= \mathbb U_0\times \cdots \times \mathbb U_{N-1} ,
\end{aligned}
\end{equation}
where $\mathbb{U} \subset \mathbb{R}^{N n_{\mathrm{u}}}$ is the set of the constrained sequence of the control inputs, $H \in \mathbb S_{++}^{Nn_{\mathrm{u}}}$ is a symmetric positive definite matrix defined as
\begin{equation}
H = \underbrace{\begin{bmatrix}
        R&  & \\
        &\ddots & \\
        &&R
\end{bmatrix}}_{\widetilde{R}} + 
\begin{bmatrix}
    \Phi_1\\\vdots \\\Phi_N
\end{bmatrix}^\top 
\underbrace{\begin{bmatrix}
        Q&  & \\
        &\ddots & \\
        &&P
\end{bmatrix}}_{\widetilde{Q}}
\underbrace{\begin{bmatrix}
        \Phi_1\\\vdots \\\Phi_N
\end{bmatrix}}_{\Phi}
\end{equation}
such that $\Phi_k = [A^{k-1}B,\cdots, AB, B, 0, ..., 0] \in\mathbb R^{n_x\times N n_u}$. Then, matrices $S\in\mathbb R^{n_{\mathrm{x}} \times N n_{\mathrm{u}} }$ and $D \in \mathbb R^{n_{\mathrm{x}} \times n_{{\mathrm{x}}}}$ are constructed as
$
S = \Psi^\top\widetilde{Q} \Phi$ and $D = \Psi^\top\widetilde{Q} \Psi
$
for $\Psi^\top = [A,\dots, A^{N}]^\top$. 

The main idea of MPC design in receding horizon policy is to solve Problem~\eqref{eq:mpc_problem_qp} within each sampling time to determine an optimal sequence of control actions $u^\star$ for a given initial condition $x_0$, and then, apply the first input $u_0^\star$ to the controlled plant. 

\begin{assumption}
\label{assumption:roa}
For the MPC problem in~\eqref{eq:mpc_design_problem}, the set of feasible initial conditions is (sub)set of the corresponding region of attraction. Technically, the terminal penalty in~\eqref{eq:mpc_design_problem} is determined to satisfy Theorem 1 in~\cite{Limon2006}.
\end{assumption}

Solving problem~\eqref{eq:mpc_problem_qp} to the optimal solution under Assumptions~\ref{assumption:1}, \ref{assumption:roa} leads to the Lyapunov descent~\cite{rawlings2017model} 
\begin{equation}
\label{eq:assymp_stability_mpc}
V(x_1) \leq V(x_0) - \|x_0\|_Q^2 ,
\end{equation}
where $x_1 = Ax_0 + Bu_0^\star$, ensuring the asymptotic stability for closed-loop systems under the receding horizon MPC control policy. Note the solution of QP in~\eqref{eq:mpc_problem_qp} leads to the $u_0^\star = f(x_0^\star)$, where $f: \mathbb{U}^{n_{\mathrm{u}}} \rightarrow \mathbb{R}^{n_{\mathrm{x}}}$ has the form of the piecewise affine (PWA) function, see~\cite{B17}.  

Dealing with~\eqref{eq:mpc_problem_qp}, using an online solver to achieve the optimal solution $u^\star$ can be computationally intractable within the given sampling time. On the other hand, the real-time implementation of iterative optimization procedures yields suboptimal solutions, as we need to stop the algorithm after reaching a certain number of iterations.
Thereafter,~\eqref{eq:assymp_stability_mpc} cannot be applied to ensure asymptotic stability in general. In the following section, we will propose a generalized stopping criterion with a fixed number of iterations for an arbitrary linearly convergent optimization algorithm to solve~\eqref{eq:mpc_problem_qp} such that an $\epsilon$-suboptimal solution results in an asymptotic stability guarantee in~\eqref{eq:assymp_stability_mpc}. 

\section{Generalized Stopping Criterion}
\label{sec:main_contribution}

We consider solving the MPC problem~\eqref{eq:mpc_problem_qp} by using the real-time implementation of the first-order optimization algorithms online. This yields a sequence of suboptimal solutions by reaching a predefined finite number of iterations $m$ during every sampling period. Such a suboptimal solution is defined as $u^m_0$ at the current time instant and is applied into the closed-loop LTI system
\begin{equation}
x_0^+ = Ax_0 + B u^m_0.
\end{equation}
To guarantee the asymptotic stability of the real-time MPC, we need to augment \eqref{eq:assymp_stability_mpc} by term $V(x_0^+)$ into the form 
\begin{equation}
\label{eq:cl_stability}
V(x_0^+) \leq V(x_0) - \left(\|x_0\|_Q^2 -V(x_0^+) +V(x_1) \right) .
\end{equation}
Enforcing~\eqref{eq:cl_stability} to hold, we need the following property of $V(\cdot)$ representing Lipschitz continuity property
\begin{equation}
\label{eq:property_V}
|V(\widetilde{x}_1)-V(\widetilde{x}_2)|\leq \eta_1 \|\widetilde{x}_1-\widetilde{x}_2\|_2 + \frac{\eta_2}{2} \|\widetilde{x}_1-\widetilde{x}_2\|_2^2
\end{equation}
providing for the properly selected real-valued constants $\eta_1 \geq 0 $ and $\eta_2 > 0$. Note,~\eqref{eq:property_V} is satisfied by design, if Assumption~\ref{assumption:1} holds.
\begin{proposition}[\!\!\cite{MAYNE2000}]
\label{proposition:1}
In the origin, such a local neighborhood $\mathcal B_r$ with radius $r$ exists, that if $\widetilde{x}_1,\widetilde{x}_2\in\mathcal B_r$ holds, then no constraints are active at the solution of~\eqref{eq:mpc_problem_qp}. 
\end{proposition}
Proposition~\ref{proposition:1} implies that for any $x\in\mathcal B_r$, the value function is reduced to $V(x) = x^\top P x$ and $\eta_1=0$ in~\eqref{eq:property_V} if $\widetilde{x}_1,\widetilde{x}_2\in\mathcal B_r$, i.e., holds
\begin{equation}
    \label{eq:remark1}
    |V(\widetilde{x}_1)-V(\widetilde{x}_2)|\leq \frac{\eta_2}{2} \|\widetilde{x}_1-\widetilde{x}_2\|_2^2.
\end{equation}
Next, we adjust \eqref{eq:property_V} to express $|V(x_0^+) -V(x_1)|$ formula complemented with applying the state update resulting in
\begin{equation}
\label{eq:lyapunov_suboptimality}
\begin{aligned}
|V(x_0^+)-V(x_1)|\leq &\eta_1 \|B(u_0^m-u_0^\star)\|_2 \\
&\qquad+ \eta_2 \|B(u_0^m-u_0^\star)\|_2^2.
\end{aligned}
\end{equation}
Substituting $u_0^m = \Gamma u^m$ and $u^\star_0 = \Gamma u^\star$ for $\Gamma = [I\;0\;...\;0] \in \mathbb{R}^{n_u\times Nn_u}$, we rewrite 
\begin{equation}
B(u_0^m-u_0^\star) = B\Gamma(u^m-u^\star)
\end{equation}
in~\eqref{eq:lyapunov_suboptimality} such that holds
\begin{equation}
\label{eq:augmented_lyapunov_suboptimality}
\begin{aligned}
|V(x_0^+)-V(x_1)|\leq &\eta_1 \|B\Gamma(u^m-u^\star)\|_2 \\
&\qquad+ \eta_2 \|B\Gamma(u^m-u^\star)\|_2^2.
\end{aligned}
\end{equation}

If we take the first-order optimization algorithm with a linear convergence rate, then the following applies
\begin{equation}
\|u^{m+1} - u^\star\|_2\leq \kappa \|u^m - u^\star\|_2,
\end{equation}
where $\kappa < 1$ represents convergence factor, $u^m$ stands for QP solution at $m$-th iteration of algorithm, and $u^\star$ is the optimal solution of QP in~\eqref{eq:mpc_problem_qp}. Consequently, for a given $m$ iterations, we have 
\begin{equation}
\label{eq:linear_convergence}
\|u^{m} - u^\star\|_2\leq \kappa^m \|u^0 - u^\star\|_2.
\end{equation}

Specifically, under the assumption of $Q\in\mathbb S_{+}^{n_{\mathrm{x}}}$, $R\in\mathbb S_{++}^{n_{\mathrm{u}}}$, and $\mathbb U$ is a compact convex set, we have two widely-used first-order method algorithms: (i) the projected gradient descent method (PGDM) and (ii) the alternating direction method of multipliers (ADMM), achieving linear convergence. 

\begin{example}[Projected Gradient Descent Method] ~\\
\label{example:PGDM}
The projected gradient descent method is an extension of the original gradient descent method by including the constraints through projection into the constraint set $\mathbb{U}$. Applying the PGDM algorithm to the MPC design problem in the form of QP~\eqref{eq:mpc_problem_qp} results in an iteration
\begin{align}
u^{m+1} := ~& \mathrm{Proj}_{\mathbb U} (u^{m} - \alpha \nabla J(u^m, x_0)) ,
\end{align}
where $\alpha > 0$ is the step size and $u^m$ is the initial guess at $m$-th iteration. We consider $J(u, x_0)$ is $\mu$-strongly convex and $L$-smooth, such that $\mu$ and $L$ are computed by evaluating the eigenvalue of $H$ in~\eqref{eq:mpc_problem_qp}. If we determine the step size $\alpha = \frac{1}{L}$, then we have linear convergence of PGDM given by
\begin{eqnarray}
\label{eq:kappa_PGDM}
\kappa = 1-\frac{\mu}{L}
\end{eqnarray}
such that $\kappa < 1$, see~\cite{Bubeck2015}.
\end{example}

\begin{example}[Alternating\,Direction\,Method\,of\,Multipliers]
\label{example:ADMM}
The alternating direction method of multipliers is an algorithm described in more detail in~\cite{Boyd2011}. Generally, the algorithm distributes the original optimization problem into smaller-scaled problems that are solved quickly. Applying ADMM to QP problem in~\eqref{eq:mpc_problem_qp} leads to following iterations
\begin{subequations}
\begin{align}
    u^{m+1} =& \min_{u} \;J(u;x_0) + u^\top \lambda^m + \frac{\rho}{2}\|u-v^m\|_2^2 ,\label{eq:local_step}\\
    v^{m+1} = & \min_{v\in\mathbb U}\;-v^\top \lambda^m + \frac{\rho}{2}\|u^{m+1}-v\|_2^2 , \\
    \lambda^{m+1} = & \lambda^m + \rho(u^{m+1}-v^{m+1}) ,
\end{align}
\end{subequations}
for some initial guesses of global coordination variable $v^m\in \mathbb R^{Nn_{\mathrm{u}}}$, 
dual variable $\lambda^m\in \mathbb R^{Nn_{\mathrm{u}}}$, and tuning parameter $\rho > 0$.
The evaluation of the linear convergence rate of ADMM for QPs refers to Section IV of~\cite{Ghadimi2015}, resulting in
\begin{equation}
\label{eq:kappa_ADMM}
\kappa = \frac{1}{2}\|2M-I\| ,
\end{equation}
where the matrix $M$ depends on the formulation of the MPC problem in~\eqref{eq:mpc_problem_qp} and has the form
\begin{equation}
\label{eq:def_m}
M = \widetilde{G} - \widetilde{G}(I+\widetilde{G})^{-1}\widetilde{G}
\end{equation}
for $\widetilde{G} = \rho \, GH^{-1}G^\top$. The matrix $G \in \mathbb{R}^{2Nn_{\mathrm{u}} \times Nn_{\mathrm{u}}}$ is determined by the matrix representation $Gu\leq w$ of input constraints from MPC problem in~\eqref{eq:mpc_problem_qp}, where vector $w \in \mathbb{R}^{2Nn_{\mathrm{u}}}$. 

\end{example}

Then, combining~\eqref{eq:augmented_lyapunov_suboptimality} with~\eqref{eq:linear_convergence} for Examples~\ref{example:PGDM},~\ref{example:ADMM}  yields inequality
\begin{align}\label{eq:lyapunov_eta}
    |V(x_0^+)-V(x_1)|\leq& \bar \eta_1 \|u^m-u^\star\|_2 + \bar \eta_2 \|u^m-u^\star\|_2^2\\\notag
    \leq &\bar \eta_1 \kappa^m \|u^0-u^\star\|_2 + \bar \eta_2 \kappa^{2m} \|u^0-u^\star\|_2^2
\end{align}
for the real-valued constants $\bar \eta_1 = \eta_1 \sqrt{\nu_{\max}(\Gamma^\top B^\top B\Gamma)}$ and $\bar \eta_2 = \frac{\eta_2}{2} \nu_{\max}(\Gamma^\top B^\top B\Gamma)$. Here, $\nu_{\max} (\cdot)$ defines the maximal eigenvalue for a given matrix. 

\begin{assumption}
\label{assumption:gamma}
We have the real-valued constant $\gamma > 0$ that for any feasible solution $x_{0}$ of~\eqref{eq:mpc_problem_qp} satisfies
\begin{equation}
    \label{eq:def_gamma}
    \|u^\star\|_2\leq \gamma \|x_0\|_Q .
\end{equation}
\end{assumption}
Since $u^\star$ is a piecewise affine function of $x_0$, the constant $\gamma$ in~\eqref{eq:def_gamma} can be determined offline, i.e., before running the real-time MPC control (see e.g.~\cite{Leung2021}). Furthermore, the initialization also satisfies $\|u^0\|_2\leq \gamma \|x_0\|_Q$ enforced by re-scale in the algorithm, if necessary. Then, \eqref{eq:lyapunov_eta} reads
\begin{equation}
\label{eq:stability_condition}
\begin{aligned}
    |V(x_0^+)-V(x_1)|\leq  2\bar \eta_1 &\gamma \kappa^m \|x_0\|_Q \\
    &\qquad+ 2\bar \eta_2 \gamma^2 \kappa^{2m}  \|x_0\|_Q^2 .
\end{aligned}
\end{equation} 

\begin{figure}[htbp!]
    \centering
    \begin{subfigure}{0.45\linewidth}
        \includegraphics[width=\linewidth]{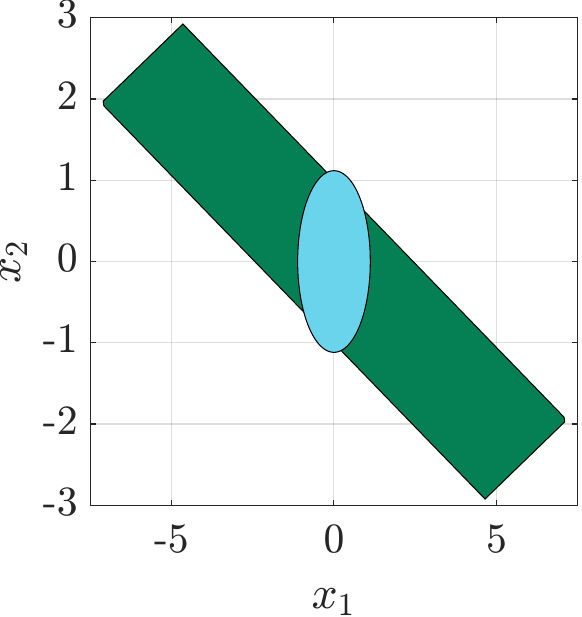}
        \caption{Assumption~\ref{assumption:3} violated.}
        \label{fig:second_sub}
    \end{subfigure}
    \hfill
    \begin{subfigure}{0.45\linewidth}
        \includegraphics[width=\linewidth]{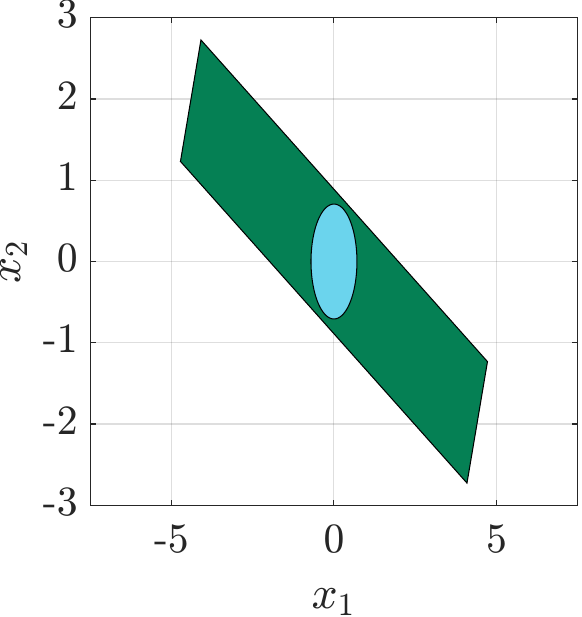}
        \caption{Assumption~\ref{assumption:3} holds.}
        \label{fig:third_sub}
    \end{subfigure}
    \hfill
    \caption{Example of set $\mathbb Q$ (blue) and LQR-based control invariant set (green) for two setups of matrix pairs $(Q,R)$ of MPC design problem \eqref{eq:mpc_design_problem}: a) $\mathbb Q$ is not the subset of the LQR-based control invariant set, b) LQR-based set contains set $\mathbb Q$.}
    \label{fig:LQR_Q_sets}
\end{figure}

\begin{assumption}
    \label{assumption:3}
    We assume that if we have non-empty convex set $\mathbb Q := \{ x_0 \in \mathbb R^{n_\mathrm{x}}  ~|~ \|x_0\|_Q \leq  1 \}$, then $\mathbb Q\subseteq \mathcal B_r $ holds.
\end{assumption}
Note that Assumption~\ref{assumption:3} can be enforced by adjusting the scaling matrices $Q$ and $R$ of MPC problem~\eqref{eq:mpc_design_problem} as shown in Figure~\ref{fig:LQR_Q_sets}. 

Consequently, a finite integer-valued parameter $\overline m$ exists such that the inequality~\eqref{eq:stability_condition} is satisfied. Finally, this finite number of iterations $\overline m$ guarantees the asymptotic stability of the real-time implementation of the MPC designed by~\eqref{eq:mpc_design_problem}.

\begin{theorem}[Generalized stopping criterion]
\label{thm:generalized_stopping_criterion}
Let Assumptions~\ref{assumption:1}--\ref{assumption:3} hold. If constant $\overline{m} \in \mathbb{N}$ satisfies 
\begin{equation}
    \label{eq:m_conservative}
    \overline{m} > \log\left(2\bar{\eta}_1\gamma+2\bar{\eta}_2\gamma^2\right)/\log\left(1/\kappa\right) ,
\end{equation}
then the control action $u_{0}:= \Gamma u^{\overline{m}}$ implemented to the LTI system~\eqref{eq:lti_system} ensures the asymptotic stability of the MPC in~\eqref{eq:mpc_problem_qp}. 
\end{theorem}

\begin{proof}
\label{proof:1}
The proof is established in two steps. First, if $ \|x_0\|_Q > 1$ holds, the upper bound on \eqref{eq:stability_condition} is given by
\begin{equation}
    \label{eq:upper_bound_FO_geq_1}
        |V(x_0^+)-V(x_1)|\leq \underbrace{\left( 2\bar \eta_1 \gamma + 2\bar \eta_2 \gamma^2 \right)\kappa^{\overline{m}}}_\beta   \|x_0\|_Q^2,
\end{equation}
for $\|x_0\|_Q^2 \geq \|x_0\|_Q$ and for $\kappa < 1$ leading to $\kappa^{\overline{m}}>\kappa^{2{\overline{m}}}$. Next, by combining~\eqref{eq:cl_stability} with~\eqref{eq:upper_bound_FO_geq_1}, we have
\begin{equation}
    V(x_0^+) \leq V(x_0) - \left((1-\beta)\|x_0\|_Q^2\right),
\end{equation}
where $(1-\beta) > 0$ is sufficient to achieve asymptotic stability. Therefore, the minimum number of iterations $\overline{m}$ is determined from following inequality
\begin{equation}
    \label{eq:m_condition_1}
    \beta = \left( 2\bar \eta_1 \gamma + 2\bar \eta_2 \gamma^2 \right) \kappa^{\overline{m}} < 1,
\end{equation}
that straightforwardly implies~\eqref{eq:m_conservative} to hold.

Secondly, if $ \|x_0\|_Q \leq 1$, using Proposition~\ref{proposition:1} and Assumption~\ref{assumption:3}, we have~\eqref{eq:stability_condition} in form
\begin{equation}
    |V(x_0^+)-V(x_1)|\leq 2\bar \eta_2 \gamma^2 \kappa^{2\overline{m}}  \|x_0\|_Q^2 \leq \beta \|x_0\|_Q^2,
\end{equation}
resulting into
\begin{equation}
    \label{eq:m_conservative_quadratic}
    \overline{m} > \frac{\log(2\bar\eta_2\gamma^2)}{2\log(1/\kappa)} .
\end{equation}
Consequently, the following inequality holds
\begin{equation}
    \overline{m} > \frac{\log\left(2\bar{\eta}_1\gamma+2\bar{\eta}_2\gamma^2\right)}{\log\left(1/\kappa\right)} > \frac{\log(2\bar\eta_2\gamma^2)}{2\log(1/\kappa)},
\end{equation}
that concludes the proof.
\end{proof}

Even though the algorithm to solve MPC problem~\eqref{eq:mpc_problem_qp} is stopped after $\overline{m}$ iterations determined by the generalized criterion in~\eqref{eq:m_conservative}, enforcing the asymptotic stability by this approach leads to the conservative control performance. The performance loss originates in~\eqref{eq:upper_bound_FO_geq_1} representing the upper bound on the original requirement formulated in~\eqref{eq:stability_condition}, i.e., represents just the necessary condition. Therefore, there may exist fewer iterations guaranteeing asymptotic stability of the closed-loop LTI system in~\eqref{eq:lti_system} under receding horizon MPC control policy. 

\section{Numerical Case Study}
\label{sec:case_study}

To analyze the properties of the proposed stopping criterion, we adopted the well-known benchmark system of the double integrator system, which has the system matrices
\begin{equation*}
A = \begin{bmatrix}
    1 & 1 \\
    0 & 0.5
\end{bmatrix},~~
B = \begin{bmatrix}
    0.5 \\
    1
\end{bmatrix},
\end{equation*}
and the model predictive controller in~\eqref{eq:mpc_problem_qp} is designed with weight matrices
\begin{equation*}
Q = \begin{bmatrix}
    1 & 0 \\
    0 & 1
\end{bmatrix},~~
P = \begin{bmatrix}
    2.367 & 1.118 \\
    1.118 & 2.588
\end{bmatrix},~~
R = 1,
\end{equation*}
where $P$ is computed as a solution to the matrix Riccati equation. Furthermore, we define the input constraints in~\eqref{eq:mpc_problem_qp} as
$\mathbb{U} = \{u \in \mathbb{R} ~|~ -1 \leq u \leq 1\}$, and finite prediction horizon as $N=10$. 

The simulations were executed in \texttt{MATLAB}. For solving the double integrator problem with ADMM-based MPC and nominal MPC, we incorporated the \texttt{QUADPROG} solver. Note that the PGDM method can be implemented without the need for any external optimization tools. Furthermore, we have designed the ADMM-based MPC algorithm in a way where local step \eqref{eq:local_step} was completely distributed in the time domain to $N$ computing units. Both algorithms disposed of two stopping criteria: either satisfying the convergence condition or meeting a maximum number of iterations.

The results for PGDM-based MPC were obtained using the following setup of the parameters $L = 3\,200$ and $\mu = 2$ leading to $\kappa_\text{PGDM} = 0.9992$ based on \eqref{eq:kappa_PGDM}. Then, the maximum number of iterations was calculated according to Theorem~\ref{thm:generalized_stopping_criterion} by~\eqref{eq:m_conservative} as $\overline{m}_\text{PGDM} = 172$ using parameters $\gamma = 1$, $\eta_1 = 0.4$, $\eta_2 =0.1$\,. The convergence condition was defined as $\|\nabla J\| \leq \varepsilon$ with $\varepsilon = 10^{-3}$.

The parameters for the ADMM-based MPC algorithm were set based on \eqref{eq:kappa_ADMM} as $\kappa_\text{ADMM} = 0.9980$ and formula from Section IV of~\cite{Ghadimi2015} as $\rho = 3.1231$. The maximum number of iterations was determined according to Theorem~\ref{thm:generalized_stopping_criterion} as $\overline{m}_\text{ADMM} = 14$ considering $\gamma = 1$, $\eta_1 = 0.2$, $\eta_2 =0.3$. Furthermore, convergence condition was specified as $\|\lambda^{l+1}-\lambda^l\|\leq\varepsilon$ for $\varepsilon=10^{-3}$.
\begin{figure}[htbp!]
\centering
\vspace{-0.4cm}
\includegraphics[width=0.95\linewidth]{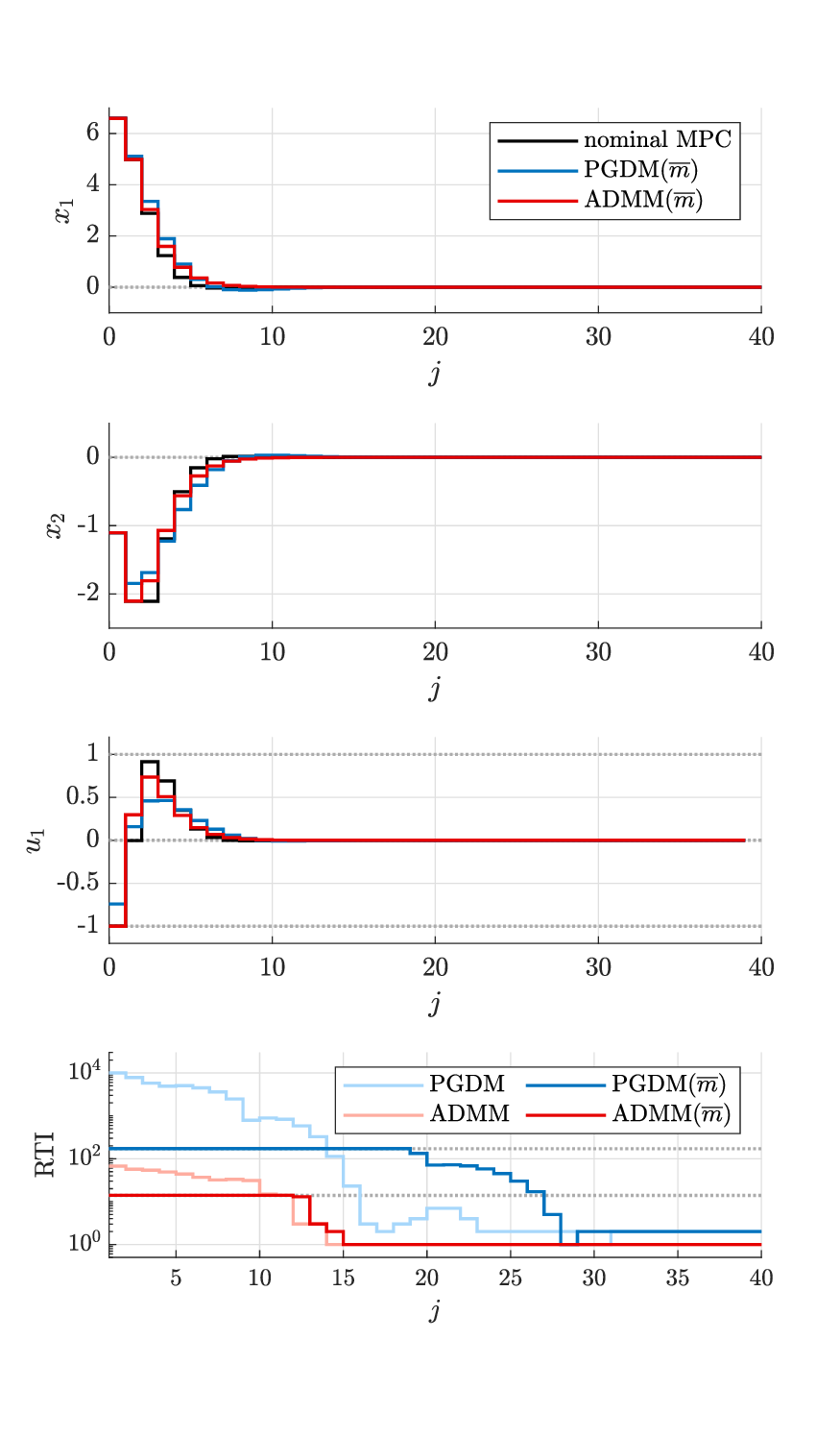}
\vspace{-1.0cm}
\caption{Control performance of the nominal MPC (black), PGDM-based MPC (blue), and ADMM-based MPC (red) after $\overline{m}$ iterations. The bottom graph depicts the number of real-time iterations (RTI) for ADMM and PGDM for unbounded stopping criterion (light blue, light red) and for predefined $\overline{m}$ (dark blue, dark red) iterations. }
\label{fig:DI_results}
\end{figure}  

The results of the numerical simulations of the closed-loop control for the proposed real-time approach are illustrated in Figure~\ref{fig:DI_results}, where we can see the comparison of control profiles for nominal MPC, $\overline{m}$-bounded ADMM-based MPC and $\overline{m}$-bounded PGDM-based MPC. To validate their accuracy, the ADMM-based and PGDM-based MPC algorithms were also evaluated for a maximum number of iterations $m = 15\,000$ (we will address it as ``unbounded''). However, these control performance results are not depicted in Figure~\ref{fig:DI_results} due to overlapping with the control profiles generated by the nominal MPC that serves as the reference profile without any performance loss. As shown in Figure~\ref{fig:DI_results}, the ADMM-based and PGDM-based MPC approaches lead to suboptimal solutions for $\overline{m}$. Nevertheless, each approach preserves the asymptotic stability and drives the system states into its origin. The last graph of Figure~\ref{fig:DI_results} shows the number of real-time iterations (RTI) per simulation step $j$ of unbounded ADMM-based and unbounded PGDM-based MPC and with $\overline{m}$-bounded (suboptimal) approaches as defined above. As visualized, the number of iterations for unbounded PGDM-based MPC is extremely higher than for $\overline{m}$-bounded PGDM-based MPC. Numerically $47\,794$ vs. $3\,620$ iterations per visualized $40$-steps simulation resulting in $92$\,\% decrease in the number of iterations. In comparison, the unbounded ADMM-based MPC executes $466$ iterations altogether while $\overline{m}$-bounded ADMM-based MPC performs $198$ iterations ($67$\,\% decrease). To conclude, even a drastically lower number of iterations yields the desired result with stability guarantees at the cost of suboptimal performance.

\begin{table}[htbp!]
\renewcommand{\arraystretch}{1.5}
\caption{Comparison of suboptimality rate and a number of iterations for unbounded and $\overline{m}$-bounded methods.}
\label{tab:results_di}
\begin{center}
    \begin{tabular}{|c||c|c|c|c|}
        \hline
        Method & $m_\text{avg}$ [--] & $m_{\max}$ [--] & $\delta_\text{avg}$ [\%] & $\delta_{\max}$ [\%]\\
        \hline \hline
        ADMM & 39 & 2\,331 & 0 & 0 \\
        ADMM($\overline{m}$) & 6 & 14 & 0.7 & 0.9 \\ \hline
        PGDM & 1\,873 & 9\,786 & 0 & 0\\
        PGDM($\overline{m}$) & 99 & 172 & 0.5 & 1.2 \\
        \hline
    \end{tabular}
\end{center}
\end{table}

\begin{figure}[htbp!]
\centering
\includegraphics[width=0.9\linewidth]{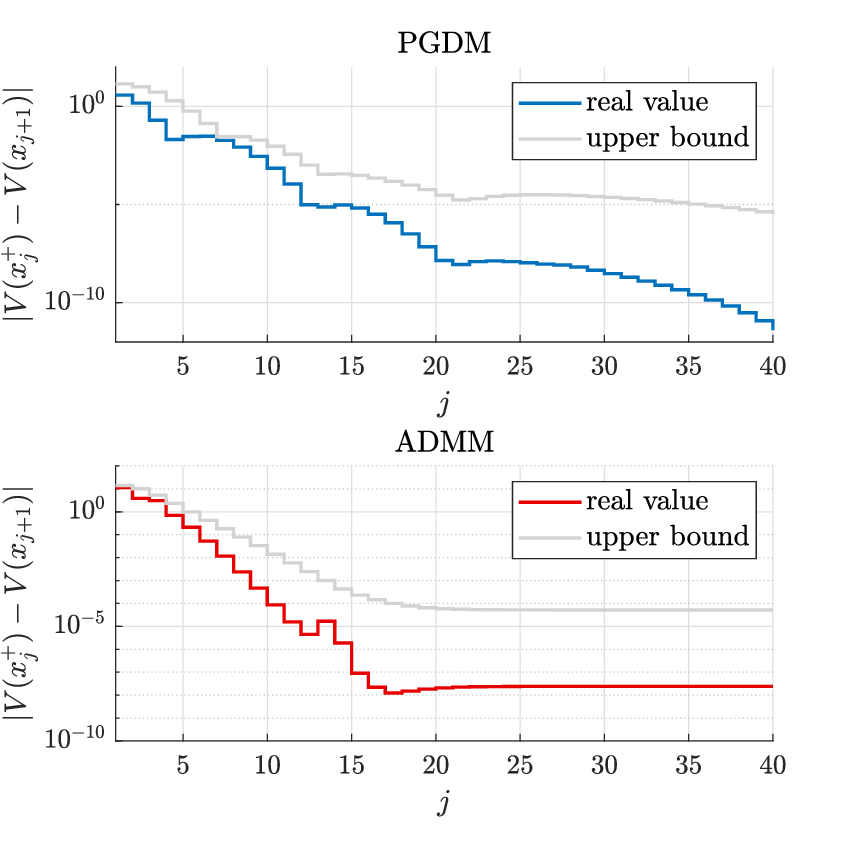}
\caption{Evolution of $|V(x_j^+)-V(x_{j+1})|$ for PGDM-based MPC (blue) and ADMM-based MPC (red) with $\overline{m}$ iterations supplemented by the respective upper bound for each method from \eqref{eq:stability_condition} (grey).}
\vspace{-0.5cm}
\label{fig:DI_conv}
\end{figure}

The stability validation of our proposed approach involves checking the inequality~\eqref{eq:stability_condition} at each simulation step. Figure~\ref{fig:DI_conv} depicts real-time $|V_j(x^+_{j})-V_j(x_j)|$ values alongside an upper bound derived as $2\bar \eta_1 \gamma \kappa^m \|x_0\|_Q + 2\bar \eta_2 \gamma^2 \kappa^{2m} \|x_0\|_Q^2$. The plot demonstrates that the real-time values of $|V_j(x^+_{j})-V_j(x_j)|$ remain under the upper bound in each step $j$.

A comparative summary is provided in Table~\ref{tab:results_di}, which analyzes suboptimality rates and iteration counts for unbounded and $\overline{m}$-bounded methods. The investigation covers control performance and evaluated iterations using metrics including the average number of iterations per simulation step $m_\text{avg}$, the maximum number of iterations per step $m_{\max}$, the average suboptimality rate per step $\delta_\text{avg}$, and the maximum suboptimality rate per step $\delta_{\max}$. The analysis involves 201 initial conditions $x_0$ representing distinct segments of the PWA control law. Suboptimality rates $\delta_j(x_0(i))$ for step $j$ and initial condition $i$ are calculated as $\delta_j(x_0(i)) = \left( |V_j(x^+_{j})-V_j(x_j)| \right) / \left( |V_{j+1}(x_1)| \right)$.

For the unbounded methods, both ADMM-based and PGDM-based MPCs exhibit zero average and maximum suboptimality rates due to $\delta_\text{avg} < 10^{-6}$ and $\delta_\text{max} < 10^{-6}$. Comparable suboptimality rates are observed for $\overline{m}$-bounded ADMM and PGDM methods. While ADMM has slightly higher suboptimality, it demonstrates lower variance than PGDM. Analysis of $m_\text{avg}$ and $m_\text{max}$ highlights that unbounded methods require substantially more iterations for optimal solutions, whereas $\overline{m}$-bounded techniques execute significantly fewer iterations (around 94\% decrease for PGDM and 84\% decrease for ADMM) while ensuring asymptotic stability.

\section{Conclusions}
\label{sec:conclusion}
This paper has presented a novel method for the real-time evaluation of linear-quadratic MPC with input constraints, offering a generalized stopping criterion representing a fixed number of iterations for first-order optimization algorithms. 
The proposed method establishes the asymptotic stability of the real-time solution. This approach significantly reduces the maximum number of iterations required for real-time evaluation of MPC while preserving the acceptable suboptimality rates. The proposed method was analyzed using a benchmark double-integrator problem, demonstrating an average reduction in the number of iterations by up to 80\,\% leading to less than 2\,\% suboptimality rate for the worst-case control scenarios. The analyzed results indicate that using a fixed number of iterations as a generalized stopping criterion has the potential to significantly decrease the run times of real-time MPC for the systems with fast dynamics and/or in the framework of the remote battery-supplied control platforms. Our future work will be focused on the elimination of two major drawbacks of our approach: (i) parameters gained by exhaustive offline numerical computations, (ii) state constraints were not considered.

\bibliographystyle{IEEEtran}
\bibliography{IEEE_bib}

\end{document}